\documentclass[12pt]{article}
\usepackage{epsfig}
\usepackage{times}

\begin{document}
\newcommand{\be}{\begin{equation}}
\newcommand{\ee}{\end{equation}}
\newcommand{\bea}{\begin{eqnarray}}
\newcommand{\eea}{\end{eqnarray}}

\begin{center}
\begin{huge}
{\bf   Model independent method for determination
of the DIS  structure of free neutron
}\end{huge} \\
\vspace{2cm}
\begin{Large}
Misak Sargsian  \\
\end{Large}
\vspace{0.6cm} Department  of Physics, Florida International University,
Miami, FL 33199,  USA  \\
\vspace{1cm}
\begin{Large}
Mark Strikman  \\
\end{Large}
\vspace{0.6cm} Department of Physics, Pennsylvania State
University, University Park, PA 16802, USA\\

\vspace{0.8cm}
\today{}
\end{center}

\begin{abstract}
\noindent 
We present a model independent procedure for extracting  deep-inelastic 
structure function of  "free'' neutron from the electron - deuteron scattering with 
protons produced in the target fragmentation region of the reaction. 
This procedure is based on  the extrapolation of $t$, which 
describes the  invariant momentum transfered to 
the proton, to the unphysical  region corresponding to the mass of the 
struck neutron. We demonstrate that the impulse approximation diagram of the 
reaction has a pole at this limit with a residue being proportional to 
the "free" neutron structure function. 
 The method is analogous to that of Chew and Low  for extraction of the 
 ``free'' pion-pion and neutron-neutron cross sections from $p(\pi,p)X$ and 
$d(n,n)pn$ reactions respectively. 
We demonstrate that in the  extrapolation  the final state interaction amplitudes are smooth 
functions of  $t$ and have negligible contribution in the extracted 
``free'' nucleon structure function.  We also estimate the range of the recoil nucleon 
momenta which could be used  for successful extrapolation procedure.
 \end{abstract}

\section{Introduction}

In spite of  three decades of studies of the partonic structure of 
nucleons  one still lacks a satisfactory knowledge of the relative $d$ and $u$ quark densities at 
large  Bjorken $x$ region.

To extract these quantities  two major approaches have been considered to 
date: one is the neutrino scattering off the  proton at large x which allows one to probe 
separately 
the $u$ and $d$ distributions. The second is the  extraction of parton distributions from 
both  protons and neutrons using inclusive scattering from the hydrogen and deuteron targets.
(In the future it would be possible also to use $W^\pm$ production at LHC\cite{LHC}.)

While deep inelastic  neutrino-proton scatterings lack  adequate statistics, 
the  inclusive electron-deuteron measurements suffer from significant nuclear effects 
an estimation of which involves the consideration of  specific models for the Fermi motion 
and the EMC  effect \cite{FS88,Rock,S93,MT}.

In our previous works \cite{MSS97,review,hix04} we outlined an alternative, model 
independent approach for  extraction of ``free'' neutron deep inelastic structure function
using semi-inclusive tagged neutron reactions, $d(e,e',p)X$, in which slow protons are detected 
in the target fragmentation region of the reaction.  These considerations were incorporated 
in  the experimental proposal which was approved recently at Jefferson Lab~(JLab)
\cite{JLab_bonus} \footnote{It is certainly of interest to study the reaction with tagged 
neutron as well which would allow a direct comparison of the scattering off a free and bound 
proton and hence identify the processes beyond the impulse approximation.}.
Preparations  for the experiment are currently under 
way and hence it is  timely to elaborate our approach quantitatively.

In our consideration of semi-inclusive deep inelastic scattering off the deuteron
 we focus in the region of relatively large
$x\ge 0.3$ were coherence length is small and hence nuclear shadowing
mechanism of the distortion of the nucleon spectrum discussed in \cite{GS}
is negligible.

The extraction procedure is based on  the observation that due to a weak 
binding in the deuteron the  singularity  of  the amplitude  in the $ t=(p_{d}-p_p)^2 $ 
-channel is much closer to the physical region  of on-shell neutron
than to  all other singularities (the closest of which would be the  
pion production threshold).
Hence one can in principle to continue analytically the scattering amplitude in $t$ and 
find the residue 
at the pole of the struck neutron propagator at  $t=m_n^2$.  
This is analogous to the Chew-Low procedure~\cite{CL} for the extraction of the pion-pion 
scattering amplitude from the pion-nucleon data.
It is worth emphasizing that in our case the analytic continuation is simpler since the
elementary amplitude does not contain factors which go through zero
at $t$ close to the  pole of the amplitude ( $t=0$ as compared to $t=m_{\pi}^2$).
Since  procedures of extrapolation work well  for the pion case, we expect them to be even 
more effective for the case of the tagged nucleon processes.

The paper is organized as follows: In section II we outline the general framework of 
semi-inclusive deep inelastic scattering off nuclear targets, concentrating on the general 
properties of the scattering amplitude in the impulse approximation as well as beyond  the 
impulse approximation.
In section III  we study the  analytic properties of  the  impulse approximation and the final state 
interaction amplitudes at kinematics of recoil proton extrapolated to 
the pole values of the struck neutron propagator. First we study the singular behavior of the 
impulse approximation amplitude in the limit $t\to m_n^2$. Then we prove the loop theorem, 
which 
states that any additional loop in the scattering amplitude removes the singularity 
associated 
with the on-shellness of the struck neutron.
In section IV we use a specific model to calculate 
the corrections to the impulse approximation due to the final state interactions.
We observe  that the deviations from the impulse approximation
due to such reinteractions is mostly concentrated at recoil angles  $\le$~100-120$^o$.
Based on the considered  final state interaction models  we elaborate the extrapolation procedure and 
demonstrate that extracted "free" neutron DIS structure function is insensitive to 
the strength of the rescattering amplitude. This explicitly demonstrates the  model 
independence of the extrapolation procedure.
We find that these procedure can be performed reliably if the tagged nucleons with momenta 
$50 \div 150$~MeV/c are detected.

\section{General Framework}
We consider semiinclusive deep inelastic scattering~(DIS) off the 
deuteron:
\begin{equation}
e + d \rightarrow e' + N + X
\label{reaction}
\end{equation}
where nucleon N is detected in the target fragmentation region.  
We define $E_e$ as the  initial energy of the electron and $E'_e$, $\theta_e$ 
as the energy and scattered angle of the final electron. $	q\equiv  (\nu, {\bf q})$ is 
the four momentum of the virtual photon, with $\nu=E_e-E'_e$ and $Q^2=-q^2$.
The recoil nucleon is described by four-momentum $p_s\equiv (E_s,{\bf p_s}$).  
We identify the masses of recoil and struck nucleons by $m_s$ and $m_N$ respectively. 
$t=(p_d-p_s)^2$ determines the invariant momentum transferred to the recoil 
nucleon, where $p_d$ is the four momentum of the deuteron.

The processes of Eq.(\ref{reaction}) with recoil proton can be used to 
extract the DIS structure  function of the neutron. Since in this case the neutron 
is bound, it requires a careful treatment of off-shell effects in 
maximally model independent way.   The approach we will discuss in this work 
is similar to one of Chew and Low~\cite{CL} who studied the issues of extracting  
$\pi^+ +\pi^0$ and $n+n\rightarrow n+n$ cross sections 
from $\pi^+ + p\rightarrow p + X$ and $n+d\rightarrow p+n+n$ reactions 
respectively. 
In their analysis they observed that the analytical structure of the  
impulse approximation amplitude ( as in Eq.(\ref{A_IA})) is such that it has a pole in 
nonphysical  region of  $t$ corresponding to the one-mass-shell kinematics of 
the  bound particles   involved in the interaction.
 
In the similar way we will analyze analytic properties of 
the scattering amplitude  of the  reaction (\ref{reaction})
in the  impulse approximation focusing on the issues related 
to the extraction of the on-shell DIS structure function  of the  neutron.

First, we summarize  the general formulae for the cross section of 
process (\ref{reaction}):
\begin{eqnarray}
& & {d\sigma \over dx dQ^2 d^3p_s/E_s}  = {4\pi\alpha_{em}^2\over x Q^4}
(1-y-{x^2y^2m_N^2\over Q^2})\times \nonumber \\
& &  \left[F_{L}^D + ({Q^2\over 2 q^2} + 
tan^2({\theta\over 2}){\nu\over m_N}F_T^D + ({Q^2\over q^2} + 
tan^2({\theta\over 2}))^{1\over 2}cos(\phi)F^D_{TL} + cos(2\phi)F^D_{TT}\right],
\end{eqnarray}
where four independent nuclear structure functions, $F^{D}_{L,T,TL,TT}$ 
depend on $Q^2,x,\alpha_s, p_{st}$, with 
$\alpha_{s} = 2{E_{s}-p^z_s \over m_D}$ being light cone momentum 
fraction of the deuteron carried out by recoil nucleon. The latter is normalized 
in such way that the sum of $\alpha_s + \alpha = 2$, where $\alpha$ is the 
similar quantity for the interacting nucleon. The Bjorken $x={Q^2\over 2m_N\nu}$ and 
$y = {\nu\over E_e}$.  The $z$ axis aligned in the direction of $\vec q$.
 In many practical considerations one integrates over the azimuthal angles $\phi$ of  
the recoil nucleon, which yields:
\begin{eqnarray}
{d\sigma\over dx dQ^2 d^3p_s/E_s} = {4\pi\alpha_{em}^2\over x Q^4}
(1-y-{x^2y^2m_n^2\over Q^2})
\left[F^{SI}_{2D} + 2tan^2({\theta\over 2})){\nu\over m_N}F^{SI}_{1D}\right],
\label{SI}
\end{eqnarray}
where: \ \  
$F^{SI}_{2D}(x,Q^2,\alpha_s, p_t) = F^D_{L} + {Q^2\over 2q^2}{\nu\over m_N}F^D_T
\ \ \ \ \mbox{and} \ \ \ \ 
F^{SI}_{1D}(x,Q^2,\alpha_s, p_t) = {F^D_{T}\over 2}$.

The theoretical description of the reaction (\ref{reaction}) at not very large values of 
$p_s< 700$~MeV/c is 
based on the assumption that it  proceeds through the interaction of 
virtual photon off one of the bound nucleons in the deuteron,
while produced particles can interact in the final state with the 
other (spectator) nucleon. Since we are interested in  $x\ge 0.3$ kinematics
it is legitimate to neglect simultaneous interaction of $\gamma^*$ with two nucleons. 
Two main diagrams will contribute to the cross section of reaction 
(\ref{reaction}): impulse approximation~(IA)~(Fig.1a) and diagram
representing a rescattering of the recoil nucleon off the products of 
DIS scattering~(Fig.1b), which we will refer as final state interaction (FSI) 
diagram.

\begin{figure}
  \includegraphics[height=.15\textheight]{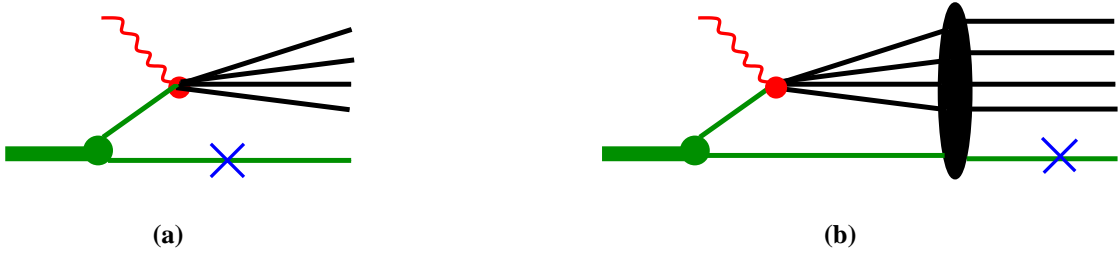}
  \caption{IA (a) and FSI (b) diagrams}
 \end{figure}

\subsection{Impulse Approximation~(IA)}
\label{IAs}

In IA the recoil nucleon is a spectator of $\gamma^*$ scattering off the bound 
nucleon $N$. One can apply Feynman diagram rules to write down the IA amplitude 
in a formal form (see e.g. \cite{FSS97,MMS}):
\begin{equation}
A^{\mu}_{IA} = <X|J^{\mu}_{em}(Q^2,\nu,p_s){p\!\!\!/_d - p\!\!\!/_s 
+ m\over m^2_N-(p_d-p_s)^2} \bar u(p_s) \Gamma_d 
= <X|J^{\mu}_{em}(Q^2,\nu,p_s){p\!\!\!/_d - p\!\!\!/_s 
+ m\over m^2_N-t} \bar u(p_s) \Gamma_d 
\label{A_IA}
\end{equation}
where  $\Gamma_d$ is the covariant $d\rightarrow pn$ transition 
vertex which is a smooth function at the pole of the struck nucleon propagator, and  
$\hat J^\mu_{em}(Q^2,\nu,x)$ represents 
the electromagnetic DIS  operator  of electron  scattering off the bound nucleon. 
Here we suppressed the 
polarization indices of the deuteron and  the nucleons. 

Taking the recoil nucleon on mass shell in Fig.1(a) and using 
$p\!\!\!/_d - p\!\!\!/_s + m \approx \sum\limits_{spins}u(p_d-p_s)\bar u(p_d-p_s)$ one can 
factorize the IA amplitude into two parts, consisting of  the DIS current of the bound 
nucleon, 
$J^{\mu}_{X,N} = <X|J^{\mu}_{em}(Q^2,\nu,p_s)u(p_d-p_s)$ and the wave function of the 
deuteron. 
The latter is expressed through the   $\Gamma_d$  vertex function and the bound nucleon 
propagator.
Since the deuteron wave function is not a  Lorentz invariant quantity  its determination 
depends on 
the reference frame in which the above factorization is  performed.

The factorization  procedure in the Lab frame of the deuteron in nonrelativistic 
limit (corresponding to the equal time quantization) yields the nonrelativistic  
deuteron wave function which is the  solution of the Schroedinger equation:
\begin{equation}
\Psi^{NR}_d(p_s) = {1\over 2\sqrt{(2\pi)^{3}E_s \ } } 
\cdot{\bar u(p_s)\bar u(p_d-p_s)\Gamma_d\over  m^2_n-(p_d-p_s)^2}.
\label{psi_nr}
\end{equation}
It is normalized as $\int |\Psi_d(p_s)|^2  d^3p_s=1$.
This correspondence usually achieved in the {\em virtual nucleon}~(VN) approximation in which 
the scattering is described in the LAB frame of the 
nucleus and electrons scatter  off  the virtual nucleon whose virtuality is 
defined by the kinematic parameters of the spectator nucleon. 
In this case  the form of the wave  function is defined through the evaluation 
of the IA amplitude at the one-mass shell pole of the spectator nucleon 
propagator in the Lab frame. This yields an off- energy - shell state of the bound nucleon.

In another approximation one can formally associate the  $\Gamma_d$  vertex with Light 
Cone deuteron wave function by considering equal light cone time $\tau = t - z$ quantization. 
In this case: 
\begin{equation}
\Psi^{LC}_d(\alpha_s,p_{st}) = {\Gamma_d\over 2\sqrt{(2\pi)^{3} \ } ({4(m^2+p^2_{st})
\over \alpha_s(2-\alpha_s)}-M_d^2)}
\label{psi_lc}
\end{equation}
normalized as $\int |\Psi_d(\alpha_s,p_{st})|^2  d^2p_{st} {d\alpha\over \alpha}=1$.
 In this approximation referred as  the {\em  light cone}~(LC) approximation~\cite{FS81} 
the scattering  is described in the light cone reference frame where 
the wave  function is evaluated at the pole of the spectator nucleon in 
the LC reference frame. This yields an off - light cone - energy  ($E+p_z$ )-shell 
state of the bound nucleon.

The above described factorization of the scattering amplitude into two parts in IA allows 
one to express the nuclear DIS structure functions  through 
the convolution of  bound nucleon DIS structure functions and the nuclear spectral function, 
$S$ as follows:~\cite{MSS97,SSS}:
\begin{eqnarray}
F^{SI}_{2D}(x,Q^2,\alpha_s,p_t) & = & {S(\alpha_s,p_t)\over n}{m_N \nu\over pq}\nonumber \\
&\times &\left[(1+cos\delta)^2(\alpha+{pq\over Q^2}\alpha_q)^2 + 
{1\over 2}sin^2\delta{p_t^2\over m_N^2}\right]F_{2N}^{eff}(\tilde x,Q^2,\alpha,p_t),\nonumber \\
F_{1D}^{SI}(x,Q^2,\alpha_s,p_t) & = &  {S(\alpha_s,p_t)\over n}
\left[F_{1N}^{eff}(\tilde x, Q^2,\alpha,p_t) + {p_t^2\over 2 pq}
F^{eff}_{2N}(\tilde x, Q^2,\alpha, p_t)\right],
\label{IA}
\end{eqnarray}
where $F_{1N}^{eff}$ and  $F_{2N}^{eff}$ are the structure functions of the bound nucleon and 
$sin^2\delta = {Q^2\over {\bf q}^2}$.
The nuclear spectral function, $S$  describes the probability of finding an interacting nucleon 
in the target with momentum ($\alpha$,  $p_t$) and 
a recoil nucleon in the final state of the reaction with momentum
($\alpha_s$, $p_{st}$). Note that in IA $\alpha_s + \alpha = 2$ and 
$\bf {p_t} = -{\bf p_{st}}$. $S$ is a model dependent quantity whose form 
depends on the framework/formalism used to describe the interaction. 

In VN approximation $n={M_d\over 2(M_d-E_S)}$ and for the spectral function one 
obtains~\cite{SSS}:
\begin{equation}
S^{IA}_{VN}(\alpha_s,p_{st})  = E_s\Psi^2_d(p_s),
\label{S_vn}
\end{equation}
where $\int S^{IA}_{VN}(\alpha_s,p_{st}{d\alpha_s\over \alpha_s}d^2p_t = 1$.
It can be shown~\cite{SSS}  that $S^{VN}$ has a normalization  such that in 
Eq.(\ref{IA}) it  conserves the number of the baryons involved in the reaction. On the other hand 
the momentum sum  rule is not satisfied. This reflects the fact that due 
to the  virtuality of the interacting nucleon  part of its  momentum is shared by 
the non-nucleonic degrees of freedom in nuclei.

In LC approximation~\cite{FS81} $n=(2-\alpha_s)$ and:
\begin{equation}
S^{IA}_{LC}(\alpha_s,p_{st}) = \Psi_{d}^{LC}(\alpha_s,p_{st}) = 
{E_k\over 2-\alpha_s}\Psi^2_d(k),\ \  
\alpha_s = {E_k-k_z\over E_k}, \ \  p_{st} =  k_t.
\label{S_lc}
\end{equation}
The remarkable property of the LC spectral function is that it fulfills both 
baryon number conservation ($\int S^{IA}_{LC}(\alpha_s,p_t){d\alpha_s\over \alpha_s}d^2p_t=1$) 
and momentum sum rule ($\int \alpha S^{IA}_{LC}(\alpha_s,p_t){d\alpha_s\over \alpha_s}d^2p_t=1$)~
(\cite{FS81})  (for recent discussion see also \cite{SSS}).

\subsection{Final State Interaction}

Calculation beyond the impulse approximation  is a very complex task in DIS. 
This includes FSI contribution in which a reinteraction is taken place between spectator 
nucleon and the products of DIS scattering off the struck nucleon.
The typical FSI amplitude can be represented as in  Fig.1b. Calculation of this 
 requires a detailed understanding of the dynamics of DIS as well as its 
hadronic structure at the final state of the reaction.  However the local character of 
DIS scattering off one of the nucleons   in the deuteron allows us to represent   
FSI diagram of Fig.1b  in the most general form as follows:
\begin{equation}
A_{FSI} = \sum\limits_{X'}\int{d^4p_{s'}\over i(2\pi)^4}
 \langle X,s|\hat A_{FSI}\cdot G(X') \cdot \hat J^{em} (Q^2,x) 
 {p\!\!\!/_d - p\!\!\!/_{s'}+m_N\over  (p_d-p_{s'})^2-m^2_N+i\epsilon}
{p\!\!\!/_{s'}+m_{N_2}\over p^2_{s'}-m^2_{N_1}+i\epsilon}\Gamma_d
\label{A_FSI}
\end{equation}
where $\hat J^{em} (Q^2,x)$ and $\hat A_{FSI}$ represent  operators 
of DIS and FSI scattering and $G(X')$ is a notation for the propagation of 
intermediate state $X'$.  The amplitude in Eq.(\ref{A_FSI}) is too general to 
allow any practical calculation. 
Our further discussion is 
based on the presence of at least one integration over the momentum of spectator nucleon in
Eq.(\ref{A_FSI}). Complexity of the intermediate state $X'$ and its final state interaction 
may lead to additional integrals over momenta in Eq.(\ref{A_FSI}). Obviously for purpose 
of demonstrating the lack of singularity it is sufficient to consider the case of 
one integration over the momentum of the spectator nucleon.

\section{Analytic Properties at the Pole}

Next we discuss the analytic properties of IA ~(Eq.(\ref{A_IA})) and FSI~(Eq.(\ref{A_FSI}))  
amplitudes at the pole of the struck nucleon propagator.  

\subsection{IA Amplitude}

We first consider IA amplitude, in which the momentum of the struck nucleon 
is fixed by the kinematics of the spectator nucleon.  It follows from Eq.(\ref{A_IA}) that 
IA amplitude has 
a pole at $t=m_N^2$ which corresponds to the struck-nucleon being on-mass shell. Using the 
following relation between $t$ and kinetic energy of the recoil nucleon, $T_s$:
\begin{equation}
t = -2M_dT_s + m_N^2 - |\epsilon_d|(M_d+m_N-m_s)
\label{tTs}
\end{equation}
it is straightforward 
to represent  the amplitude of Eq.(\ref{A_IA})  in the following form:
\begin{eqnarray}
A_{IA} =  <X|J^{em}(Q^2,x)|n>\bar u(p_d - p_s)\bar u(p_s) {\Gamma_d\over 
|\epsilon_d|(M_d + m_N - m_s) + 2 M_d T_s}.
\label{A_IA2}
\end{eqnarray}
This equation shows that the  pole is associated  with the negative value of 
kinetic energy of the spectator nucleon, equal to the half of the magnitude of the deuteron 
binding energy $\epsilon_d$:
\begin{equation}
T_s^{pole} = -{|\epsilon_B|\over 2}(1 + {m_n-m_p\over M_d}) \approx 
-{|\epsilon_d|\over 2}.
\label{pole}
\end{equation}
 Despite being  in nonphysical region of the reaction (\ref{reaction})  the very fact of the 
existence of the pole indicates that the extrapolation of the measured semiinclusive cross 
section to the positive region of $t$ (negative 
kinetic energy region of the spectator nucleon)  will allow us to isolate the ``free'' cross 
section of  the electron - nucleon scattering. This can be true only if the singularity is 
unique to the IA amplitude.

\subsection{Loop Theorem and non IA Amplitudes }

Now we will prove  that since all FSI type amplitudes contain at least one loop integral in the 
amplitude,  they are regular at the one-shell  pole of the struck nucleon propagator.

It is sufficient to prove
that no singularities exist for the amplitude that contains at least  one loop 
involving the spectator nucleon, to prove that all higher order diagrams will 
have a smooth behavior at $t\rightarrow m_N^2$ limit. 
In considering the FSI diagram, using Eq.(\ref{tTs}),  we will analyze the analytic 
behavior of $A_{FSI}$ with respect to $t$.
Due to the relation in Eq.~(\ref{tTs}) all we need to demonstrate is  
that $A_{FSI}$ is a smooth function in  $T_s\rightarrow -{|\epsilon|\over 2}$ limit. 

In Eq.(\ref{A_FSI}) we first 
evaluate the integral over $d^0p_{s'}$ at  the pole of the  spectator nucleon:\\ 
$\int dp_{s'}^0 {1\over p^2_{s'} - m^2_N + i\epsilon} = -{i 2\pi\over 2E_{s'}}$, then 
introducing $k = p_s - p_{s'}$ one obtains:
\begin{eqnarray}
A_{FSI} = \int{d^3k\over (2\pi)^3}\left[
 {{\langle X,s|\hat A_{FSI}}G(X')\hat J^{em} (Q^2,x) |N\rangle 
\bar u(p_d-p_s-k)\bar u(p_{s}-k)\over 2(m_N+T_s-k_0)} \right. \nonumber \\
\left. \times{\Gamma_d\over -M_d^2 + 2M_d(m+T_s-k_0)}\right]. 
\label{A_FSI2}
\end{eqnarray}
To estimate  the contribution of this amplitude at the pole of the struck nucleon 
propagator it is enough now to  estimate the integral 
\begin{equation}
\int{d^3k\over  M_d|\epsilon_b|+ 2M_d T_s-2M_dk_0}.
\label{int}
\end{equation}
For relatively small $k$,
using the relation:  $k_0 = \sqrt{m_N^2+p^2_s} - \sqrt{m^2_N+(p_s-k)^2} \approx
{\vec p_s \vec k\over m_N} - {k^2\over 2m_N}$ we see that in the limit
$T_s\rightarrow  -{|\epsilon_b|\over 2}$ integrand in Eq.(\ref{int}) is finite,
\begin{equation}
{k^2dk\over  2M_d({k^2\over 2m_N})}\rightarrow {dk\over 2}. 
\label{fsipole}
\end{equation}
This demonstrates that the integral in Eq.(\ref{A_FSI2})
is finite in  $T_s\rightarrow -{|\epsilon_B|\over 2}$ limit . 
Note that  the integral  converges also  in the  large $k$ limit since in this case the  
integrand  is  suppressed due to the amplitude of rescattering of the system $X$ off 
the spectator nucleon.
This result demonstrates that all contributions that contain at least one 
rescattering with spectator nucleon have a smooth behavior at the pole 
of the struck nucleon propagator as compared to the singular behavior of 
IA amplitude in Eq.(\ref{A_IA2}).
 
\section{Extracting ``Free'' Neutron DIS Structure Function}
 
We demonstrate the procedure of extraction of ``free'' neutron structure functions using 
specific models for evaluation of the  FSI. We then demonstrate that the 
result of the extrapolation is practically insensitive to the parameters which define the 
strength of FSI in these models. 
The latter indicates the model independence of  the procedure.

\subsection{Model for FSI}
We discuss a model in which the FSI  of the  spectator nucleon occurs coherently 
with the system  produced in the  deep inelastic electron- bound nucleon 
scattering~(see also \cite{CKK}). 
Such approach is rather 
well founded for energies relevant to Jefferson Lab\cite{JLab_bonus}, in which the 
multiplicity of produced system  is  restricted to few hadrons.  
Due to the same reason a nonspectator contribution in which a nucleon with small
momentum is produced in the $\gamma^*+N\rightarrow N+X$ process is strongly suppressed 
at $x\ge 0.3$ as it requires scattering off a nucleon with a very large momentum 
$\alpha > 1+x$.
Therefore contributions to the cross section associated with the incoherent rescattering 
will not contain any terms singular at $t\rightarrow m_N^2$.
Thus we expect that the main 
contribution to DIS cross section will come from the diagrams of type of Fig.1b  in which 
FSI amplitude could be estimated in analogy to the processes of  $e + d \to e+p+n$ considered in 
Ref.\cite{FSS97,MMS,FGMSS}. In this approach one arrives at the similar eikonal form of the 
rescattering 
amplitude treating the effective  cross section and the t-dependence of the rescattering as  free 
parameters. 
In the calculation we neglect a small effect of modification of 
Fermi momentum of the struck nucleon in the $\gamma^*N $ amplitude due to FSI.  This is justified 
due to 
the fact  that  the dominant process of rescattering corresponds to the conserved light-cone 
fraction of the interacting nucleon\cite{FSS97,MMS} (see also below).  Within this approximation  
we can  factorize $\gamma^* N$ DIS amplitude 
from the FSI. This allows us to use the whole theoretical 
framework described in Sec.\ref{IAs} by replacing $S^{IA}$  with the 
distorted spectral functions, $S^{DWIA}$ which include both  IA and  FSI effects.

Using the procedure of the generalized eikonal approximation of Ref.\cite{FSS97,MMS} within 
virtual nucleon approximation for the distorted  spectral function one obtains:
\begin{eqnarray}
S^{DWIA}_{VN}  & =  & E_s\left[\Psi^2_d(p_s) -  \right.  \nonumber  \\
& &  {1\over 2}\sqrt{{M_d-E_s\over E_s}}Im\int {d^2k_{t}\over (2\pi)^2} f(k_t)
\left[\Psi_d^\dagger(p_s)\Psi_d(\tilde p_s) - i\Psi_d^\dagger(p_s)\Psi^\prime_d(\tilde p_s) 
\right]+ \nonumber \\
& & \left.  \left({1\over 4}\sqrt{{M_d-E_s\over E_s}}\right)^2
\left|\int {d^2k_{t}\over (2\pi)^2} f(k_t)
[\Psi_d(\tilde p_s) - i \Psi^\prime(\tilde p_s)]\right|^2\right],
\label{SFSI_vn}
\end{eqnarray}
where $\tilde p_s \equiv (\tilde p_{sz},\tilde p_{st})= (p_{sz}-\Delta_{VN},p_{st}-k_t)$ and $
\Delta_{VN} = {(M_d+\nu)\over {\bf q}}(E_s-m) + {W^2-W^2_0\over 2 {\bf q}}$. 
Here $W^2 = (q+M_d-p_s)^2$ and 
$W^2_0 = (q+m)^2$.  $\Psi^{\prime}_d$ results from the non-pole contribution in the integration 
over the longitudinal component  of the transferred momentum 
(for details  and  for the expression of $\Psi^{\prime}_d$ see Appendix B of Ref.\cite{MMS}).
Similar to the IA case an averaging over  the polarizations of the  deuteron is assumed.

In LC approximation~\cite{inprogress} for distorted spectral function we obtain:
 \begin{eqnarray}
S^{DWIA}_{LC}  & = & {E_k\over \alpha}\left[\Psi^2_d(k) -  \right.  \nonumber \\ 
& &  {1\over 2}\sqrt{{m\over E_k}}Im\int {d^2l_{t}\over (2\pi)^2} f(l_t)
\left[\Psi_d^\dagger(k)\Psi_d(\tilde k) - i\Psi_d^\dagger(k)\Psi^\prime_d(\tilde k) \right]+  
\nonumber \\
& & \left.  {1\over 16}{m\over E_k}
\left|\int {d^2l_{t}\over (2\pi)^2} f(l_t)
[\Psi_d(\tilde k) - i \Psi^\prime(\tilde k)]\right|^2\right]
\label{SFSI_lc}
\end{eqnarray}
where $\tilde k \equiv (\tilde k_{z},\tilde k_{t})= (k_{z}{m\over E_k}-\Delta_{lc},k_t-l_t)$ and 
$\Delta_{lc} =  {M_d+q_-\over  q_+}({k^2_t-\tilde k_t^2\over m\alpha_s}) + {W^2-W^2_0\over  q_+}$. 
Here $\tilde k$ represents the relative momentum of the target nucleons in the light 
cone reference frame. They are expressed through the light cone momentum fraction
and the transverse momentum of the target nucleons 
according the relations given at the RHS of Eq.(\ref{S_lc}).

It follows from Eqs.(\ref{SFSI_vn}) and (\ref{SFSI_lc}) that  in both cases the distorted 
spectral function has three distinctive terms. The first term represents the IA term 
discussed already in 
Sec. \ref{IAs}, the second term represents the interference between IA and FSI amplitudes 
which screens the $\gamma^*d$ cross section,  while 
the third term is the pure rescattering contribution. 

Note that one of the important differences between VN and LC approximations, is that at 
high energy limit (${\bf q}, q_0 \gg m$), $\Delta_{VN}$ is finite approaching to $(E_s-m)$, while
$ \Delta_{LC}$ vanishes. This reflects the fact that at high energy limit the light cone momentum 
fraction of the interacting nucleons (such us $\alpha_s$) conserves during 
final state interaction.

The amplitude $f$ in both Eqs.(\ref{SFSI_vn}) and (\ref{SFSI_lc})  describes the rescattering 
of the spectator nucleon off the products of deep inelastic scattering off struck nucleon.  
Since characteristic momentum transfer in FSI is relatively small ($\ll m$) we can model  
this amplitude in the form of: $f(k) = \sigma_{eff}(i+\alpha)e^{-{B\over 2}k_\perp^2}$ 
with $\sigma_{eff}$,  $\alpha$ and  $B$ being free parameters.  
For our numerical estimates we fix $B=8$~GeV$^2$ and $\alpha=-0.2$ corresponding 
to characteristic values of hadronic interaction at small momentum transfer.
We will allow variation of $\sigma_{eff}$ between zero and  60--80~mb. The latter numbers 
represents that characteristic cross sections of  scattering of the spectator nucleon
off the ``X'' state containing one baryon   and up to two mesons.

The distorted spectral functions of Eqs.(\ref{SFSI_vn}) and (\ref{SFSI_lc})  now can be  used in 
Eq.(\ref{IA}) to estimate the $F^{SI}_{2D}$ and $F^{SI}_{1D}$  in the approximation which
includes both IA and FSI effects. Such approximation we will refer  as distorted wave impulse 
approximation~(DWIA).

In Fig.2 we represent the predictions of VN and LC models for $F^{SI}_{2D}$ calculated within 
PWIA and DWIA  approximations. 
One can see from the figure that if the strength of $XN$ interaction is varied within 
a reasonable range ($\sigma_{XN}\le 60$~mb) one finds a rather 
broad range of predictions. This highlights the necessity of the 
dedicated measurements~\cite{deeps,Klimenko} of 
semiinclusive DIS reaction off the deuteron covering wide kinematic range of the recoil nucleon 
that will allow the focused 
comparison of the predictions of different theoretical approaches. 
It is important however that at $\alpha=1$  the  difference
between numerical results of VN and LC  approaches 
is negligible leaving the source of the main  uncertainty to  the strength of the FSI.
 
\begin{figure}[t]

\centerline{\epsfig{height=9cm,width=9cm,file=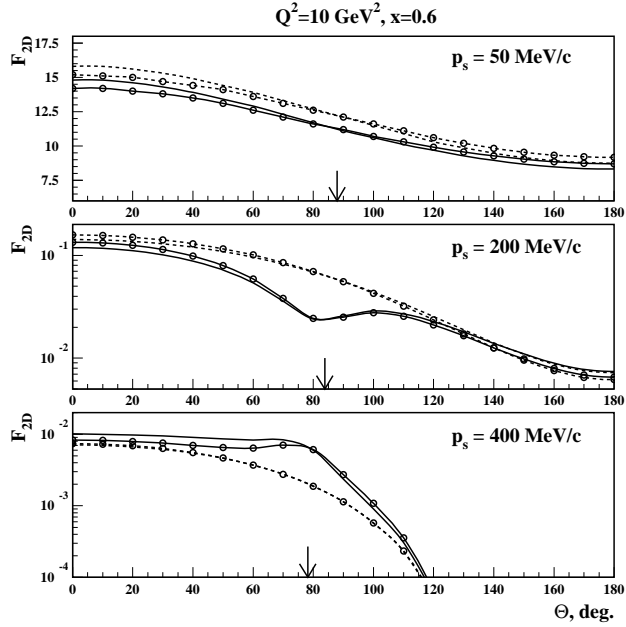}}
  \caption{Angular dependence of $F^{SI}_{2D}$ for different values of recoil nucleon momenta. 
Dashed and solid curves 
correspond to PWIA and DWIA predictions  within VN approximation. Curved labeled by circles 
corresponds to LC 
predictions. The total cross section  of $XN$ state in the FSI amplitude is taken as $60~$mb. 
Arrows identify the 
angles corresponding to $\alpha=1$.}
 \end{figure}

\subsection{Extraction Procedure for $F_2$}

We discuss now the procedure that allows us to extract the ``free'' $F_{2}$ for the struck 
nucleon. 
One starts with 
the assumption that we don't know the strenght of the FSI for which we can assume 
various values of effective reinteraction cross section from $0$ to  $80$mb (see discussion 
above).   

To employ the extraction procedure based on the pole extrapolation method  we introduce an extraction
factor $I(p_s,t)$ defined as follows:
\begin{equation}
I(p_s,t) = {1\over E_s} {(m_N^2-t)^2\over [Res (\Psi_d(T_{pole}))]^2}
\cdot {1\over {m_N \nu\over pq}\left[(1+cos\delta)^2(\alpha+{pq\over Q^2}\alpha_q)^2 + 
{1\over 2}sin^2\delta{p_t^2\over m_N^2}\right]},
\end{equation}
where $Res (\Psi_d(T_{pole})) = {C\over \sqrt{2} \ \pi 2m_N}$~GeV$^{-{1\over 2}})$ 
where $m_N$ corresponds to the mass of the struck nucleon.
Here $C$ slightly depends on the particular potential used to calculate the 
deuteron wave function. For example  $C=0.3939$ for the wave function with Paris 
potential\cite{Paris} and  
$C=0.3930$ for the Bonn potential\cite{Bonn}.

Using $I(p_s,t)$ we define the extracted structure function as follows:
\begin{equation}
F^{extr}_{2N}(Q^2,x,t) = I(p_s,t)\cdot F^{SI,EXP}_{2D}(x,q^2,\alpha_s, p_t), 
\label{f2extr}
\end{equation}
where $F_{2D}^{SI,EXP}$ is the experimentally measured value of the integrated structure function 
defined in Eq.(\ref{SI}).

If only PWIA contributions were important in the cross section of the reaction (\ref{reaction}) then 
one can estimate  $F^{SI,EXP}_{2D}(x,q^2,\alpha_s, p_t)$ using Eq.(\ref{IA}) with  spectral 
functions  defined  in  Eqs.(\ref{S_vn}) and (\ref{S_lc}). In this case one observes that 
by  extrapolating   $t$ to   the value  of $m_N^2$,  $F^{extr}_{2N}$ approaches to 
$F^{eff}_{2N}(x,Q^2,\alpha=1,p_t=0) = F^{free}_{2N}(x,Q^2)$.

In the case when FSI effects are considered   we evaluate $F^{SI,EXP}_{2D}(x,q^2,\alpha_s, p_t)$
within  DWIA framework, in which  $F^{SI}_{2D,EXP}$ is 
estimated  using Eq.(\ref{IA}) with  the distorted spectral functions defined in 
Eqs.(\ref{SFSI_vn}) and (\ref{SFSI_lc}). 
In this case we again extrapolate $F^{SI,EXP}_{2D}(x,q^2,\alpha_s, p_t)$  to the values of 
$t\rightarrow m_N^2$.

To establish the appropriate extrapolation procedure one has to investigate the analytic properties 
of Eq.(\ref{f2extr}) 
with respect to the variable of $t^{\prime} \equiv t-m_N^2$.
Here one observes that  $F^{extr}_{2N}$,  at small 
$|t^{\prime}| \ll m_N^2$,  
is an explicit quadratic function of  $t^{\prime}$ with  IA-FSI interference term being proportional
to $t^{\prime}$, while the double scattering term ($\sim |A_{FSI}|^2$) is proportional to 
$t^{\prime 2}$. 

Note however  that  the $|A_{IA}|^2$ term has a (hidden) weak polynomial dependence  on 
$t^{\prime}$ due to 
higher mass singularities in the deuteron wave function 
($\psi_d \sim \sum\limits_{i} {C_i\over -t^\prime +M^2_i}$ 
with $M^2_i = (i-1)4\gamma m_0+ 2(i-1)^2m^0$ (see e.g. Ref.\cite{Bonn}, 
where $\gamma = 0.04568$~GeV and $m_0=0.17759$~GeV.  
Since the second nearest pole corresponds to rather large positive values for 
$t^\prime$ ($\approx$~0.1~Gev$^2$), based on 
Eq.(\ref{tTs}) on estimates that by restricting recoil nucleon momenta $p_s\le 250$~MeV/c 
the $|A_{IA}|^2$  term can be represented again as a quadratic function of 
$t^\prime$.  Therefore if spectator nucleon kinetic energy  (or $t^\prime$)
is much less than the energy scale corresponding to the higher mass singularity in the deuteron 
wave function,  Eq.(\ref{f2extr}) will represent a quadratic function of $t^\prime$. 
 
The above discussion  suggests that the pole extrapolation can be achieved by 
the square fit  of the experimental data measured at small and finite values of $t^{\prime}$.  
However such extrapolation can be  meaningful  only if
we can exclude the variation of $F^{extr}_{2N}(Q^2,x,\alpha,t)$ due to the change of the other 
kinematic variables involved in the reaction.  One significant variation is the $\alpha$ 
dependence of $F^{extr}$ due to the combination of ${x\over \alpha}$ entering 
in the bound nucleon structure function. This may significantly alter $t^{\prime}$ dependence 
at large $x$ kinematics. Another important effect in high $x$ kinematics is  the higher twist 
effects due to sensitivity of structure functions on  the produced final mass of the DIS 
scattering, $W_N$ at intermediate $Q^2$. Since $W^2_N\sim \alpha W^2_{N0}$ where 
$W_{N0}$ is the final mass produced off the stationary nucleon, the $W_N$ sensitivity 
will result in  additional $\alpha$ dependence of $F^{extr}$.  
Thus, in the large $x$ kinematics, especially at intermediate $Q^2$, 
the $t^{\prime}$-dependence of 
$F^{extr}_{2N}$ will change strongly with $\alpha$.

Since 
\begin{equation}
\alpha\sim 1+{p_s\cdot cos(\theta_s)\over m},
\label{alpha_theta}
\end{equation}
the  discussed above sensitivity will result in the sensitivity of the extrapolation procedure 
to the direction of the momentum of  the recoil nucleon. 
 To illustrate this point we calculate the ratio,
\begin{equation}
R = {F^{extr}_{2N}(Q^2,x,\alpha,t)\over F^{free}_{2N}(Q^2,x)}
\label{ratio_x}
\end{equation}
and study its variation with 
$t^\prime$ at different  fixed values of $\theta_s$.   Figs.3(a) and (b) demonstrate such  
calculations within 
both DWIA and PWIA using VN approximation at $x=0.7$ for  $Q^2=5$~Gev$^2$~(a) 
and $Q^2=10$~Gev$^2$~(b) kinematics.  These calculations  reveal  significant violation 
of quadratic dependence of 
$F^{extr}$ at $\theta_s>(<)~90^0$  in which case  the corresponding $\alpha$ is away from 
unity.  This reflects the fact that 
both $F^{extr}_{2N}(Q^2,x,\alpha,t)\sim F^{bound}_{2N}(Q^2,{x\over 2-\alpha},W_N,t)$ and 
$F^{free}_{2N}(Q^2,x,W_{N0})$ are sensitive functions of  $x$ and $W$ and a little mismatch  
between $x$ and $\tilde x \approx {x\over 2-\alpha}$  as well as $W_N$ and $W_{N0}$  can result 
to a substantial irregularity in $t^\prime$ dependence of $R$.  
Note that $W$ sensitivity is especially strong at small $\theta_s$ (see $\theta_s = 30^0$ curves 
in Fig.3a)  which corresponds to small (resonating)  values of $W_N$. However 
we can see from  Fig.3(b)  that the $W$ dependence diminishes with an increase of $Q^2$ due to 
suppression of the higher twist effects. 
 
Although both Fig.3a and 3.b demonstrate  substantial angular ($\alpha$) dependencies, 
they also indicate the way to  avoid these complications.  Due to relation  (\ref{alpha_theta})
 these  irregularities are gone at $\theta_s=90^0$, since in this case $\alpha\approx 1$ 
and $x\approx \tilde x$ and $W^2_N\approx W^2_{N0}$.
 Therefore one concludes that  $\alpha_s=\alpha=1$ represents the most suitable kinematical 
condition for $t^\prime$ extrapolation of $F^{extr}_{2N}(Q^2,x,t)$. It is worth mentioning that 
according to Fig.2,  $\alpha_s=1$ also diminishes 
the discrepancy  between VN and LC approaches.

\begin{figure}[t]
\centerline{\epsfig{height=10cm,width=12cm,file=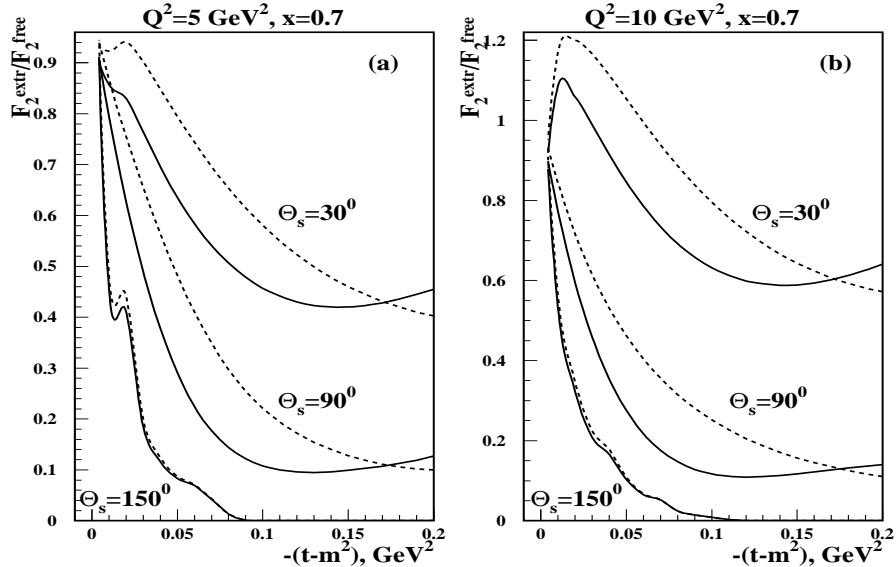}}
\vspace{-1.2cm}
  \caption{The $-(t-m_n^2)$ dependence of $R$ for different values of recoil nucleon angle. 
Dashed and solid curves 
correspond to PWIA and DWIA predictions  within VN approximation. The total cross section  of reinteraction  of 
$XN$ state in the FSI amplitude is taken  $60~$mb.}
 \end{figure}

\begin{figure}[t]
\centerline{\epsfig{height=9cm,width=9cm,file=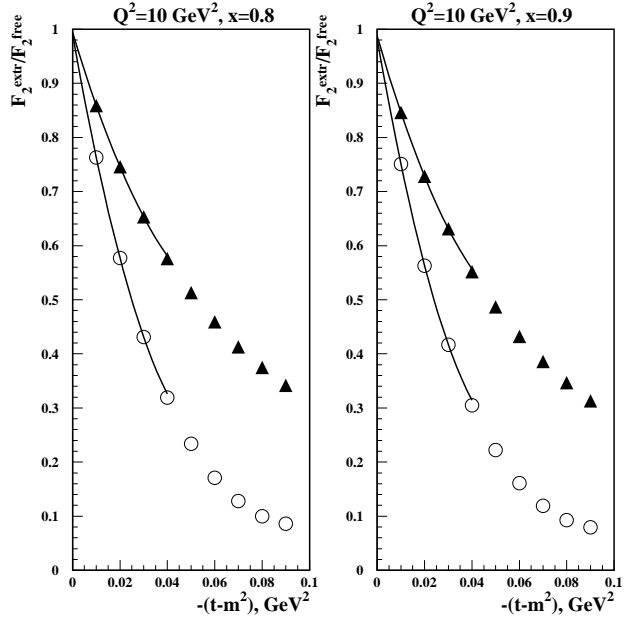}}
  \caption{Extrapolation of  $R$ at $-(t-m_n^2)\rightarrow 0$ based on the square fit using first 
three points of calculated  $R$. The triangles and open circles correspond
to $R$ (Eq.(\ref{ratio_x}) calculated within PWIA and DWIA using VN model.}
 \end{figure}

Fig.4 represents one example of the extrapolating procedure for $\alpha_s=1$ kinematics.  In this 
case $R$ is calculated using both PWIA and DWIA models within VN approach for  
$x=0.8$ and $x=0.9$ at $Q^2=10$~GeV$^2$ and for physically accessible values of $t^\prime~<~0$. 
Then we use three calculated points of $R$ to 
make a quadratic fit which is then extrapolated to the $t^\prime \rightarrow 0$ values. 
For DWIA the FSI strength is set to $80$~mb which may be considered 
somewhat above the  highest possible value 
for the cross section of the $X$-state scattering off the spectator nucleon. 
For the fitting points we choose three $R$ values calculated at $-t^\prime$ (0.1, 0.2, 0.3) 
corresponding to (54MeV/c,  89MeV/c, 114MeV/c) values of recoil proton momenta. The choice of 
these three values is motivated by the kinematic acceptance of the experiment of 
Ref.\cite{JLab_bonus}.

The quadratic function fitted to above points yields the following extrapolated values for $R$ 
at $-t^\prime = 0$: (0.9923 and  0.9888) for (PWIA and DWIA) predictions at $x=0.8$ and  
(0.9868, 0.9828) for $x=0.9$.  
These numbers demonstrate that extrapolation procedure practically eliminates the uncertainty 
due to the FSI, making FSI effects  less than $0.5\%$.  

Similar estimations within LC approximation yield for extrapolated values of $R$  
at $-t^\prime = 0$:  (0.9983, 0.9946) for (PWIA, DWIA) predictions at $x=0.8$ and 
(1.0078,1.0036) for $x=0.9$.  These numbers combined with the 
numbers obtained above in the VN approximation demonstrate that 
the overall uncertainty of extrapolation procedure using the fitting points consistent 
with the kinematics of experiment of Ref.\cite{JLab_bonus} is  of the  order of $1\%$. This 
is an unprecedented accuracy that can be achieved in extraction of high $x$  DIS 
structure function of the nucleons.

Note that due to the 
higher order singularities in the deuteron wave function, moving the fitting 
points away from the $t^\prime =0$ limit will worsen the convergence of the 
quadratically fitted function of $R$ to the unity. In this case 
an additional improvement of the procedure can be achieved if we additionally 
normalize $R$ in Eq.(\ref{ratio_x}) by the momentum distribution 
of the deuteron.

\section{Conclusions}
We have considered the pole extrapolation procedure  aimed 
at  model independent extraction of high $x$ DIS structure function of struck nucleon 
in semiinclusive DIS scattering off the deuteron.
For analysis of the validity of this procedure we considered two theoretical 
approaches (virtual nucleon and light cone) for description of semiinclusive deep 
inelastic scattering from the deuteron.  Using these approaches we modeled the final state 
interaction based on distorted wave impulse approximation. 

Within this framework we demonstrated that a simple quadratic fit  is well suited for 
on-mass shell  extrapolation of the  bound nucleon structure function. 
Although the strength of FSI is varied in the 
wide range of the scattering cross sections $0\div 80$~mb the 
overall uncertainty due to FSI and the choice of the particular (VN or LC) theoretical
approximation   is estimated to be on the level of $1\%$. This result indicates  the 
model independent character of the mass-shell extrapolation procedure.

The numerical  estimates are done for $t^\prime$ values characteristic to the experiment of 
Ref.\cite{JLab_bonus}. Thus one can expect the same level of uncertainty in the extraction 
of on-shell structure functions of nucleon if the pole extrapolation procedure is applied 
to the actual data.

\medskip
\medskip

\noindent{\bf Acknowledgments:}\\
This work is supported  by  DOE grants under contract DE-FG02-01ER-41172 and DE-FG02-93ER40771 as 
well as by the Israel-USA Binational Science Foundation Grant.

\end{document}